\documentclass{article}
\usepackage{spconf}
\usepackage{cite}
\usepackage{amsmath,amssymb,amsfonts}
\usepackage{algorithmic}
\usepackage{graphicx}
\usepackage{textcomp}
\usepackage{xcolor}
\usepackage{subcaption}
\usepackage{multirow}
\usepackage{booktabs}   
\usepackage{placeins}   
\usepackage[utf8]{inputenc}   
\usepackage{pifont}           
\usepackage{newunicodechar}   

\newunicodechar{✓}{\ding{51}} 
\newunicodechar{✗}{\ding{55}} 
\begin{document}

\title{REFESS-QI: REFERENCE-FREE EVALUATION FOR SPEECH SEPARATION WITH JOINT QUALITY AND INTELLIGIBILITY SCORING}


\name{
\begin{tabular}{c}
Ari Frummer$^{\dagger \star}$,
Helin Wang$^{\dagger}$,
Tianyu Cao$^{\dagger}$,
Adi Arbel$^{\star}$,
Yuval Sieradzki$^{\star}$
\\, 
Oren Gal$^{\ddagger}$,
Jes\'us Villalba$^{\dagger}$,
Thomas Thebaud$^{\dagger}$,
Najim Dehak$^{\dagger}$
\end{tabular}}

\address{
    $\dagger$ Johns Hopkins University, USA \\
    $\star$ Technion, Israel Institute of Technology, Israel \\
    $\ddagger$ University of Haifa, Israel \\
}

\maketitle
\begin{abstract}
Source separation is a crucial pre-processing step for various speech processing tasks, such as automatic speech recognition (ASR). 
Traditionally, the evaluation metrics for speech separation rely on the matched reference audios and corresponding transcriptions to assess audio quality and intelligibility.
However, they cannot be used to evaluate real-world mixtures for which no reference exists.
This paper introduces a text-free reference-free evaluation framework based on self-supervised learning (SSL) representations. 
The proposed framework utilize the mixture and separated tracks to predict jointly audio quality, through the Scale Invariant Signal to Noise Ratio (SI-SNR) metric, and speech intelligibility through the Word Error Rate (WER) metric. 
We conducted experiments on the WHAMR! dataset, which shows a WER estimation with a mean absolute error (MAE) of 17\% and a Pearson correlation coefficient (PCC) of 0.77; and SI-SNR estimation with an MAE of 1.38 and PCC of 0.95.
We further demonstrate the robustness of our estimator by using various SSL representations.

\end{abstract}


\vspace{-3mm}
\section{Introduction}
\vspace{-3mm}
Humans can concentrate on a particular voice or speech source in noisy surroundings. This phenomenon is known as the 'cocktail party effect'\cite{b1}. In speech processing, the corresponding challenge is accurately separating different sound sources from mixed audio signals, a task referred to as speech separation. Speech separation is typically used as a pre-processing step for speech recognition, as it helps to enhance recognition accuracy \cite{b2}.
In recent years, deep learning models have emerged as highly effective solutions for speech separation \cite{b3,b4,b5}, showcasing considerable advancements in performance compared to conventional techniques\cite{b6}.

The performance of traditional speech separation is typically evaluated using signal-based metrics, such as the SI-SNR.
In addition to the signal level metrics, task-oriented evaluations are widely employed to verify that the separated signals remain useful for downstream applications, such as WER for ASR.

Previous techniques have been proposed to estimate SI-SNR in speech separation and serve as baselines in this study \cite{b7} \cite{b8}. In addition, unsupervised quality estimation for speech enhancement has been explored~\cite{b9}. This method evaluates a single output of a general enhancement system. Furthermore, a method for predicting the WER for general speech was introduced~\cite{b10}. However, this method does not specifically predict WER based on learned relationships between representations of noisy audio and enhanced outputs. To the best of our knowledge, this work is the first to predict WER specifically in the context of speech separation.

This paper introduces a novel reference-free model designed to estimate various speech separation metrics called ReFESS-QI, utilizing the Hugging Face Transformers framework for unsupervised evaluation. The network processes audio features from both the mixed input and the separated sources to predict metrics such as SI-SNR and WER in a 3-dimensional forward pass. This model outperforms existing SI-SNR estimators and achieves comparable WER baselines without relying on text hypotheses obtained from an ASR model. 

\noindent\textbf{The key contributions of this paper are as follows}: We introduce the first text-free, reference-free WER estimator
        for speech-separated signals, We leverage self-supervised learning representations (SSLR) to create an improved SI-SNR estimator compared to existing baselines, We generalize our findings by training a set of joint estimators that predict WER and SISNR, indicating that this method can predict joint quality and Intelligibility scores.
\vspace{-5mm}

\section{Related Works}
\vspace{-3mm}
\subsection{Speech Separation}
In speech separation, the goal is to extract each speaker's utterance into a dedicated output channel. 
Given a mixture
\[
y \in \mathbb{R}^{N}, \qquad 
y = \sum_{k=1}^{K} s_k + n
\]
We seek to estimate the individual source signals.
\[
s_k \in \mathbb{R}^{N}, \qquad k = 1,\dots,K
\]
where \(n \in \mathbb{R}^{N}\) represents background noise and \(N\) is the number of time samples.  
This paper focuses on the two-speaker scenario, i.e., \(K = 2\).


Several models have been proposed to address the phenomenon of speech separation, utilizing diverse architectures and techniques. These include DPRNN, which leverages recurrent neural networks for long-sequence modeling, TasNet, based on an encoder–decoder framework, and SepFormer, which utilizes transformers\cite{b3, b4, b5}.



\vspace{-3mm}
\subsection{Speech Separation Metrics}
Speech-separation performance can be evaluated with generic signal-based metrics, such as the \textit{Signal-to-Noise Ratio (SI-SNR)} and the \textit{Perceptual Evaluation of Speech Quality (PESQ)}\cite{b11}, or with specialized downstream application-related metrics, such as the \textit{WER} for ASR.

Signal-based metrics are computed directly from the waveform by comparing with a reference. SI-SNR is estimated directly from the signal while PESQ is model-based, comparing the clean and processed signals with the aid of a built-in perceptual model to estimate perceived speech quality. 

By opposition, the \textit{WER} metric is measured for a given ASR system, by comparing the output of the ASR with a text reference, where the proportion of insertions, deletions, and substitutions required to align the recognized text with the reference quantifies the impact of separation on recognition performance.

\vspace{-3mm}
\subsection{Metrics estimators}
Here we go through three related metric estimators baselines.
In the SI-SNR baseline \cite{b8}, an approach utilizing SSLR models to predict SI-SNR was introduced. This method concatenated temporally the mixture and the two separate audio tracks before feeding them in to the SSLR's feature extractor. In contrast, our method extracts features from each audio track separately and learns the relationships between the representations in each time bin before making predictions about the metric. Moreover, the baseline is dependent on fine-tuning of the entire SSL model, which consists of millions of parameters, whereas our method does not require full fine-tuning. This allows us to preserve the knowledge from unsupervised training while only requiring light training.

The WER baseline \cite{b9} is a method that utilizes both audio and text features from the ASR hypothesis. These features are fed into the MLP head that predicts a WER estimate for a single audio.
This cascade approach presents efficiency and accuracy issues. The chaining of the results of the ASR model to the text embedding model is one point of failure when the ASR input is not accurate, and the inference of the three models is considered a heavy task.

Uni-VERSA \cite{b9} is a unified network that simultaneously predicts various objective metrics. This framework offers an efficient assessment of speech enhancement across multiple evaluation metrics. 
However, Uni-VERSA has been optimized for speech enhancement, considering only one audio channel, which is not adapted to the needs of a speech separation task,  containing by definition 2+ output channels.
\begin{figure*}[!ht]               
  \centering
  \vspace{-5mm}
  \includegraphics[width=\textwidth,]{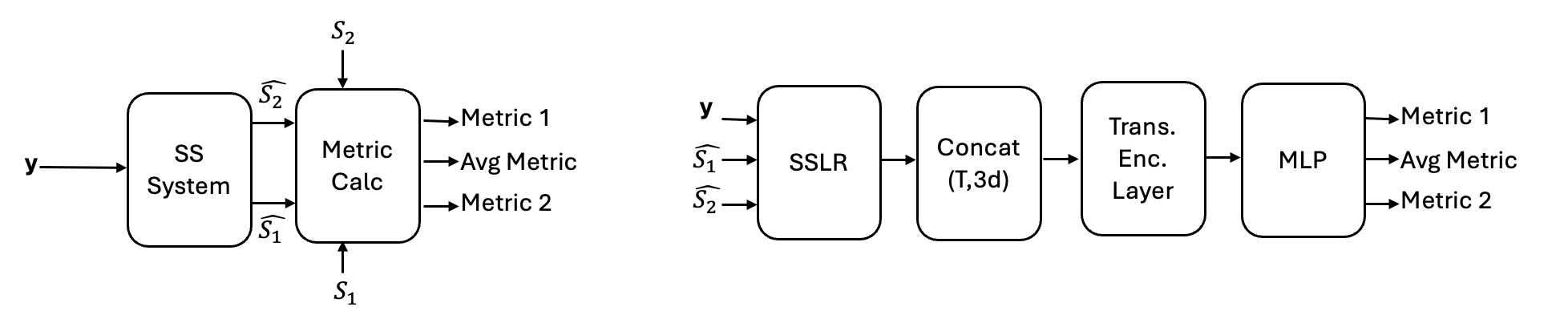}
  \caption{Left: Dataset Building Method. Right: Metric-estimator model's architecture, based on Semi-Supervised Learning Representations (SSLR), followed by a concatenation, a Transformer encoder layer, and a linear layer. 
}
  \label{fig:extractor}
\end{figure*}

\begin{table*}[!t]
  \centering                    
  \footnotesize
  \vspace{-3mm}
  \caption{Distribution of the data-set splits of the WHAMR! separated
           tracks.  ASR generated with Whisper large-V3.}
  \label{tab:joint}
  \vspace{-3mm}
  \setlength{\tabcolsep}{5pt}

  \resizebox{0.9\textwidth}{!}{      
  \begin{tabular}{l c c c c c c c }
    \toprule
    \textbf{Dataset} &
    \textbf{Split} &
    \textbf{Total Dur. (h)} &
    \textbf{Avg Dur. (S)} &
    \textbf{\#Seg} &
    \textbf{Avg \#WRD} &
    \textbf{Avg WER} &
    \textbf{Std. Dev of WER} \\ \midrule
    REAL-M & Test  &    1.3    &   5.17 & 924    &  11.2 & 49.99\%& 31.17\% \\
    WHAMR! & Test  &    8.2    &    10.7&2,750    &  16.36 & 49.36\%& 20.38\% \\
    WHAMR! & Train &     402.7   &    10.36 &140,000    &   16.41   &  49.84\%   &   31.05\%     
    \\ \bottomrule
  \end{tabular}}
\end{table*}


\vspace{-3mm}
\section{Methods}
\vspace{-3mm}
This section describes our automatic WER estimator method, as well as its extensions as an SI-SNR estimator and a joint estimator, applied to speech separation.
First, we used multiple speech separation systems to extract pairs of separated audios from mixtures.
Second, we compute the metrics from the extracted audios. The SI-SNR is computed directly and the WER from Whisper V3-Large~\cite{b12} predictions, using the Nemo toolkit\cite{b13} for text normalization.
Third, the metric estimator model is trained on the triplets of audio (mixture and 2 outputs) to predict the corresponding metrics.
\vspace{-3mm}
\subsection{Speech separation systems}
To diversify our training set, three different source separation models are used: TasNet\cite{b3} using espnet\cite{b14}, DPRNN\cite{b4}, and SepFormer\cite{b5} using speechbrain \cite{b15}.
Each separator is trained on the reverberant WHAMR! Dataset\cite{b16} to perform two-speaker separation and de-reverberation. 

\vspace{-3mm}
\subsection{Metric estimator Architecture}
\label{subsec:AA}
The estimator model performs multi-output regression to estimate a target metric, depending on the training task. 
It uses a SSL encoder~\cite{b17, b18, b19} to extract $T$ features of dimension $D_{SSL}$ from the mixture $Y$ and the two separated tracks $\hat{S_1}$ and $\hat{S_2}$, then concatenates the features along the feature axis and outputs an array of size $(T, 3\times D_{SSL})$.
Then, the array is processed by a transformer encoder layer, mean pooled across the time dimension, followed by a linear layer, to predict three continuous values: metric value for source 1, source 2, and their average. 
The model is trained with a mean squared error (MSE) loss and is evaluated using MAE, and Pearson's correlation. 
This architecture supports reference-free estimation for any signal-level metrics or downstream model value.
\vspace{-3mm}
\subsection{Metric estimator training setup}
During training, the entire network undergoes a 10,000 step warm-up period. Following the warm-up, two different learning rate schedulers are implemented to prevent overfitting. The Self-Supervised Learning model is trained using a linearly decreasing learning rate, starting at 1e-5. Meanwhile, the parameters that are trained from scratch begin with a learning rate of 1e-4 after the warm-up stage, following the same linear decreasing schedule. An Adam optimizer is used, with a batch size of 12.
\vspace{-3mm}
\section{Experiments}
\vspace{-3mm}
\label{sec:exp}

\subsection{Datasets}
We use the reverberant max version of WHAMR! \cite{b16}  which is a far-field speech separation corpus designed to simulate real-world noisy and reverberant conditions. Each mixture comprises two overlapping speakers with additive environmental noise and reverberation applied through simulated room impulse responses. Additionaly, we use 
REAL-M\cite{b7}, a real-life speech source separation dataset for two-speaker mixtures.

The WHAMR! train and validation splits were used to train and validate both the speech separation models and the metric estimators. The performances of our models were evaluated on the WHAMR! test split and the Real M dataset

We use the SS systems to create a dataset of Metrics. To ensure a balanced label distribution, we selected three checkpoints from the beginning, middle, and end epochs for dataset generation. ASR was calculated using Whisper V3-Large\cite{b12}, and text normalization was carried out with the NeMo toolkit\cite{b13}. Additionally, the  SISNR was computed using only the reference audio.
The distribution of the dataset is described in Table I and Figure 2.

\begin{figure}[!ht]
    \centering
    \includegraphics[width=1.0\columnwidth]{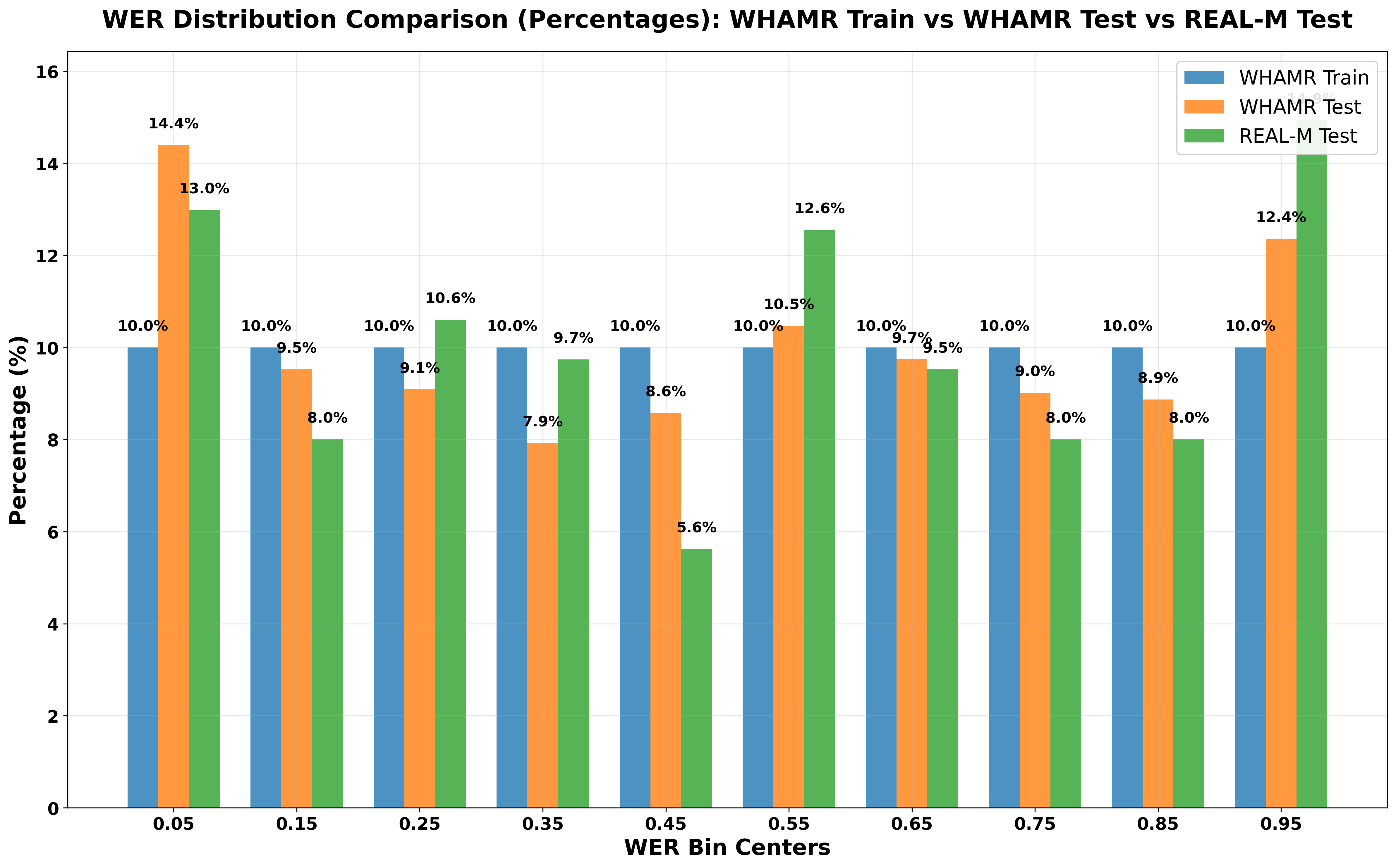}
    \caption{Average WER BIN distribution (percentage) for each mixture across dataset splits}
    \label{fig:example}
\end{figure}
\vspace{-10mm}
\subsection{Metric Estimators Training}

\subsubsection{Single metric estimators training}

For each sample, the input to the estimator comprised three-channel audio, consisting of the mixture signal along with the two separated sources. Metric scores, computed using ground-truth references, were employed as regression targets. Initially, estimators were trained separately for the SI-SNR and WER prediction tasks, following the same architecture and data processing pipeline outlined in Section \ref{subsec:AA}.
\vspace{-1mm}
\subsubsection{Joint estimators training}

In another experiment, we jointly estimated SI-SNR, and WER. To stabilize the training process and ensure consistent scaling across data samples, all labels were min-max normalized to fall within the range of $[0, 1]$. This normalization was applied globally, utilizing the minimum and maximum values observed in both the training and validation sets.

\vspace{-3mm}
\subsubsection{Ablation Studies}

Our ablation study considered three state-of-the-art self-supervised speech models as feature extractors for metric estimation: Wav2Vec 2.0\cite{b17}, WavLM\cite{b18}, and HuBERT\cite{b19}. We conduct experiments with a trainable and frozen FE.

\subsection{Generalization to New Data}
\vspace{-2.5mm}
To assess generalization, we evaluated the trained WER estimator on the REAL-M data set \cite{b7}, which is a real mixture data set that contains reverberation and noise. REAL-M is for evaluation only, and therefore will be separated using our pretrained models on WHAMR!.



\begin{table}[ht]
  \centering
  \vspace{-1mm}
  \footnotesize
  \caption{Pearson’s correlation (PCC) and mean-absolute error (MAE) for the single and average WER estimators, for multiple Feature Extractors (FE) with FE trainable or frozen, tested on the WHAMR! dataset. 
           }
  \label{tab:wer}
\resizebox{1.\linewidth}{!}{
  \begin{tabular}{l c c c c c c c c}
    \toprule
    \textbf{Models} & \textbf{FE} & \textbf{FE TR} 
                   & \textbf{Test set}
                    & \textbf{Head} & \textbf{PCC} & \textbf{MAE} &\textbf{A.PCC} &\textbf{A.MAE} \\ \midrule
    \cite{b10}                 & HuBERT+RoBERTa& ✗ & \multirow{6}{*}{WHAMR!}  & MLP            & .746 & \textbf{.15} & -&- \\
    \textbf{Ours}            & W2V2& ✓      & &Tr. + Lin. & .762 & 0.19 & .808 & 0.17 \\
    \textbf{Ours}            & Hubert& ✓    & &Tr. + Lin. & \textbf{.775} & .17 &\textbf{.824} &\textbf{0.14} \\
    \textbf{Ours}            & Hubert& ✗    & &Tr. + Lin. & .701 & .21 &.755 & 0.17 \\
    \textbf{Ours}            & WavLM& ✓     & & Tr. + Lin.& .775 & .17 &.823&0.14\\
    \textbf{Joint}           & Hubert& ✓    & &Tr. + Lin. & .769 & .17 &.819&0.14\\ 
    \midrule
    \cite{b10}                 & HuBERT+RoBERTa& ✗ & \multirow{3}{*}{REAL-M}  & MLP            & .413 & .28 &-&-\\
    \cmidrule{5-7}
    \textbf{Ours}            & Hubert& ✓    &   & \multirow{2}{*}{Tr. + Lin.}            & .547 & \textbf{.22} &.510&0.22\\
    \textbf{Joint}           & Hubert & ✓   &  &           & \textbf{.548} & .25 &\textbf{.513} & \textbf{0.21}\\ 
    \bottomrule
  \end{tabular}
  }
  \vspace{-3mm}
\end{table}
\vspace{-5mm}
\section{Results}
\vspace{-3mm}
\subsection{WER estimator}
In terms of WER estimation, Table \ref{tab:wer} shows that our method that lightly trains the FE achieves better results than the baseline without using the Text hypothesis. Our method achieves an MAE of 0.17 and a PCC of 0.775 for the single metrics estimation, and an MAE of 0.14 and PCC of 0.824 for the average metrics estimation, while the baseline achieves an MAE of 0.15 and PCC of 0.746 using an additional ASR model and text embedder for the evaluation. Moreover, our estimation method with freezing the FE achieves competitive results to the baseline, which also uses frozen models, but more pretrained models in inference. We achieve our best results by utilizing Hubert representations. In the generalization experiment using new data from the Real-M dataset, our method exceeded the baseline across all metrics.

\vspace{-3mm}
\subsection{SI-SNR estimator}
For the SI-SNR estimation task, Table \ref{tab:sisnr} shows that our method performs better than the SOTA baseline. For all the SSLRs, our methods perform better in terms of PCC and MAE. 
Even when using frozen SSL representations, our method outperforms the baseline, which was using a pre-trained SSL extractor, on both MAE and PCC.
Our best performances are obtained using WavLM representations. 

\vspace{-3mm}
\subsection{Joint estimators}
The experiments in Table \ref{tab:joint} show that this method can predict more than one metric. In terms of WER, that joint estimator achieves similar results than the single metrics estimator, both on the WHAMR! test set and the REAL-M datasets. 

For the out-of-domain evaluation (REAL-M), the WER estimator achieves an MAE of 0.25 and a PCC of 0.548.



\vspace{-2mm}
\section{Conclusion}
\vspace{-3mm}

This paper presents a novel text-free reference-free model designed to estimate separately two metrics: WER and SI-SNR from the outputs of a speech separation system.
We proposed a new architecture leveraging self-supervised learning representations that outperformed significantly previous baselines without text nor audio references.
We craft our own training dataset from WHAMR! and three different speech separation systems, designed to offer a balanced distribution of WER and SISNR metrics, then train and evaluate our proposed framework on it.
The proposed framework demonstrates state of the art performances with different SSL representations, with and without finetuning of the encoder.
Furthermore, we show our model can generalize to real mixtures when evaluated on the REAL-M dataset, and can work to estimate jointly both the WER and the SI-SNR.

\begin{table}[t!]
  \centering
  \footnotesize

  \caption{Pearson’s
           correlation (PCC) and mean-absolute error (MAE) for single and average SI-SNR estimation, for multiple Feature Extractors (FE), with FE trainable or frozen, tested on the WHAMR! dataset.
           }
  \label{tab:sisnr}
    \resizebox{1.0\linewidth}{!}{
  \setlength{\tabcolsep}{5pt}
  \begin{tabular}{l c c c c c c c}
    \toprule
    \textbf{Model} & \textbf{FE} & \textbf{FE TR} 
                   & \textbf{Head} & \textbf{PCC} & \textbf{MAE} & \textbf{A.PCC} & \textbf{A.MAE}  \\
    \midrule

    Baseline\cite{b8} & W2V2  & ✓ & \multirow{2}{*}{MLP}           & .817 & 2.01  & .916 & 1.70\\
    Baseline\cite{b8} & WavLM  & ✓ &                    & .823 & 1.616 & .922 & \textbf{1.14}\\
    Baseline\cite{b8} & HuBERT & ✓ &                    & .823 & 1.937 & .922 &1.63 \\
    \cmidrule{4-6}
    \textbf{Ours}   & W2V2   & ✓ & \multirow{4}{*}{Tr. + Lin.}    & .947  & 1.475 &.952 & 1.44 \\
    \textbf{Ours}   & WavLM  & ✓ &     & \textbf{.951}  & \textbf{1.388} &\textbf{.956} &1.34 \\
    \textbf{Ours}   & HuBERT & ✓ &     & .949  & 1.474 &.955 & 1.446 \\
    \textbf{Ours}   & HuBERT & ✗ &     & .936  & 1.459 & .943 & 1.400 \\
    \bottomrule
  \end{tabular}
  }
  \vspace{-2mm}
\end{table}

This work has certain limitations, on the speech separation systems, the evaluation data and the applications.
First, our proposed model used the same dataset to train as the speech separation models (the WHAMR! set), and the same set of speech separation models was used to generate the training and the testing data, limiting the potential generalization on new speech separation systems. 
Second, we adopted a uniform distribution of the scores in the training data, which may not be best suited to align with the out-of-domain data, and we use a unique model to extract all transcripts for the WER evaluation (Whisper).
Finally, the proposed metric estimator is based on self-supervised learning (SSL) systems, which typically requires large GPUs for inference, thus it will not be easy to use during training, only for evaluation.

Future work could explore the possibility of incorporating this estimator as a tool to guide unsupervised speech separation techniques during their training or their evaluation.
To improve the usability of this technique, we will use pruning and distillation techniques to reduce the size of the estimator, making it more fit for end to end training of larger speech separation systems without drastically increasing the required size of GPUs.

\vspace{-5mm}







\bibliographystyle{IEEEbib}
\bibliography{refs}

\end{document}